# KIMAP: Key-Insulated Mutual Authentication Protocol for RFID

Atsuko Miyaji and Mohammad ShahriarRahman

School of Information Science, Japan Advanced Institute of Science and Technology
1-1 Asahidai, Nomi, Ishikawa, Japan 923-1292,
Email: {miyaji,mohammad}@jaist.ac.jp

## Abstract:

Radio-Frequency IDentification (RFID) is an automatic identification method, relying on storing and remotely retrieving data using devices called RFID tags or transponders. An RFID tag is an object that can be applied to or incorporated into a product, animal, or person for the purpose of identification using radio waves. These RFID tags are heavily constrained in computational and storage capabilities, and raise numerous privacy concerns in everyday life due to their vulnerability to different attacks. Both forward security and backward security are required to maintain the privacy of a tag i.e., exposure of a tag's secret key should not reveal the past or future secret keys of the tag. We envisage the need for a formal model for backward security for RFID protocol designs in shared key settings, since the RFID tags are too resource-constrained to support public key settings. However, there has not been much research on backward security for shared key environment since Serge Vaudenay in his Asiacrypt 2007 paper showed that perfect backward security is impossible to achieve without public key settings. We propose* a Key-Insulated Mutual Authentication Protocol for shared key environment, KIMAP, which minimizes the damage caused by secret key exposure using insulated keys. Even if a tag's secret key is exposed during an authentication session, forward security and `restricted' backward security of the tag are preserved under our assumptions. The notion of `restricted' backward security is that the adversary misses the protocol transcripts which are needed to update the compromised secret key. Although our definition does not capture perfect backward security, it is still suitable for effective implementation as the tags are highly mobile in practice. We also provide a formal security model of KIMAP. Our scheme is more efficient than previous proposals from the viewpoint of computational requirements.

## 1 Introduction

Passive Radio Frequency Identification (RFID) tags used in stores and industries are getting ubiquitous nowadays. These RFID tags are heavily constrained in computational and storage capabilities, and this has made the RFID tags vulnerable to different attacks. This vulnerability raises numerous privacy concerns in everyday life. One of the main issues of RFID security and privacy has to do with malicious tracking of RFID-equipped objects. While tracking RFID tags is typically one of the key features and goals of a legitimate RFID system, unauthorized tracking of RFID tags is viewed as a major privacy threat. Both forward and backward securities are required to maintain the privacy of the tag. Forward security means that even if the adversary acquires the secret data stored in a tag, the tag cannot be traced back using previously known messages [2, 12]. Backward security means the opposite, i.e., even if the adversary acquires the secret data stored in a tag, the tag cannot be traced using subsequently known messages. In other words, exposure of a tag's secret should not reveal any secret information regarding the past or the future of the tag. Moreover, indistinguishability means that the values emitted by one tag should not be distinguishable from the values emitted by other tags [12, 19].

### 1.1 Related Work

Many privacy-preserving mutual RFID authentication schemes have been proposed in recent years [1, 6, 7, 9, 14, 15, 26, 18, 24, 34]. MSW (Molner,Soppera,Wagner) protocols [18] use a hierarchical tree-based keying to allow tag identification/authentication. But MSW protocols have a security aw whereby an adversary who compromises one tag is able to track/identify other tags that belong to the same families (tree branches) as the compromised tag [1]. Hash-based Access Control (HAC), as defined by [34], is a scheme which involves locking a tag using a one-way hash function. The scheme allows a tag to be tracked, because the same secret key is used repeatedly [7]. An authentication protocol for RFID from EPCGlobal Class-1 Gen-2 standards was introduced by [7]. Both the authentication key and the access key are updated after a successful session in order to provide forward security. However, [26] showed that [7] is not backward and forward-secure, because an attacker that compromises a tag can identify a tag's past interactions from the previous communications and the fixed EPC of the tag, and can also read the tag's future transactions. Another lightweight mutual RFID authentication was proposed by [15] which focused mainly on achieving indistinguishability and forward security features. However, this protocol does not support backward security, neither did it came up with any formal definition of indistinguishability or forward security. There are also some other privacy-preserving RFID protocols that address untraceability and forward security [6, 9, 24]. However, all these protocols have the same drawback, that is, they cannot provide backward security. LK and SM schemes [14, 26] have recently described RFID authentication schemes satisfying both forward and backward security. However, [26] has been shown to be vulnerable to an attack where an adversary breaks the forward security [25]. The scheme proposed in [14] cannot provide backward security if the current secret key is compromised [20]. Again, [16] proposed an RFID authentication protocol, and defined a formal model for backward security assuming that a part of the secret key is leaked only – not the whole secret key. In other words, it is assumed that for backward security, the time available to the adversary is limited, because in most cases, output of RFID tag is valid only during a certain time period. It follows that the adversary can obtain a limited amount of leaked information from side-channel attack that leads to partial leakage of the secret key in the tag. Such an assumption of partial leakage is not realistic in practice for RFID tags, since there can be various strong attacking devices that can fully compromise the whole secret key of a resource constrained RFID tag. Backward security has also been studied in the mutual authentication protocol proposed by [36] assuming that the adversary can passively attack the protocol transcript. The notion of backward security in that work is such that the secret key and the tag-specific id cannot be compromised by an adversary. In other words, an adversary is not allowed to break in to the tag's memory. That means, some kind of tamper-resistance has to be assumed for the tag. Such a restriction is not practical since today's attacking devices are strong enough to attack a tag's memory and today's low cost tags are not tamper-resistant.

**Key Insulated Public Key Cryptosystem:** In Key Insulated Public Key Cryptosystem [10], a 'master secret' key SK* is stored in a device which is physically secure and hence resistant to compromise. All decryption, however, is done on an insecure device for which key exposure is expected to be a problem.

The lifetime of the protocol is divided into distinct periods 1,...,N. At the beginning of each period, the user interacts with the secure device to derive a temporary secret key which will be used to decrypt messages sent during that period; we denote by $SK_i$ the temporary key for period i. Ciphertexts are now labeled with the time period during which they were encrypted. The insecure device, which does all actual decryption, is vulnerable to repeated key exposures. The goal is to minimize the effect such compromises will have. The notion of security is that the adversary will be unable to determine any information about messages sent during all time periods other than that in which a compromise occurred. However, this approach is not suitable for RFID environment. RFID tags do not have the capability to communicate with the server by themselves, and do not have the resources for computing public key encryption which requires high computation ability. Low-cost RFID tags usually are capable of performing only the hash function, random number generation and bitwise XOR operations. Therefore, we need to come up with a key-insulated mechanism that works for RFID environment.

**Key Insulation in Public-Channel:** In [3], Bellare et al. show that it is impossible to achieve public-channel key insulated security in the face of an active adversary (who has full control of the channel). That is, in a model which allows an interactive channel-update protocol and evolving channel-security keys, an active adversary can always succeed in breaking the scheme. The reason is that after the adversary breaks in, it obtains the user's channel-security key and can thus impersonate the user. Authentication (such as an authenticated session key exchange) does not prevent this, since the adversary acquires all the user's credentials via the break-in. This negative result is particularly strong because their public-channel key insulated model is as generous as one can get, while keeping the spirit of key insulation. Bellare et al. show that public-channel key insulated protocol is possible against an adversary that is allowed only a passive attack on the communication channel. (Meaning it can eavesdrop.) This result is significant because, it shows that key insulation is at least possible over a channel where the adversary may be able to eavesdrop but finds it hard to inject or corrupt transmissions. However, assuming passive adversary in case of RFID is not practical, as it is easy for an adversary to break into a tag's memory. We thus need to make a different assumption that is practical for RFID environment, where the adversary is active but has limited effect after key-compromise due to the use of key-insulation mechanism.

### 1.2 Motivation:

Since the adversary is able to trace the target tag at least during the authentication immediately following compromise of the tag secret, perfect backward security makes no sense. Therefore, a minimum restriction should be imposed to achieve backward security, such that the adversary misses the necessary protocol transcripts to update the compromised key. Although this assumption for backward security is true for certain classes of privacy-preserving RFID protocols (i.e., for shared key environment), it is clearly not true for some other cases. For instance, Vaudenay shows an RFID protocol based on public-key cryptography that is resistant to this attack [30]. The same result was shown by [21] for mutual authentication protocols. However, our notion of backward security is true for privacy-preserving RFID protocols based on shared secrets that are updated on each interaction between tag and reader, which is the focus of this paper. Backward security is thus harder to achieve than forward security in general, particularly under the very constrained environment of RFID tags. However, backward security is never less important than forward security in RFID systems. In the case of target tracing, it suffices to somehow steal the tag secret of a target and collect interaction messages to trace the future behaviors of the particular target. Without backward security, this kind of target tracing is trivial. In the case of supply chain management systems, even a catastrophic scenario may take place without backward security: if tag secrets are leaked at some point of tag deployment or during their time in the environment, then all such tags can be traced afterwards. We thus envisage the need for a formal model for backward security in RFID protocol designs (even if not perfect) in addition to the well-recognized forward security.

### 1.3 Our Contribution

We propose KIMAP, a privacy-preserving Key Insulated Mutual RFID Authentication Protocol for shared key environment which provides both forward and `restricted' backward security through key insulation. Even if a tag's secret key is exposed during an authentication session, forward security and `restricted' backward security of the tag are preserved under our assumptions. The notion of `restricted' backward security is that the adversary misses the protocol transcripts needed to update the compromised secret key. The protocol also provides indistinguishability between the responses of tags in order to provide privacy of a tag. We also provide a formal security model to design our privacy-preserving protocol. Our assumptions for indistinguishability and forward/restricted backward security are similar to the assumptions made in previous work.

**Organization of the Paper:** The remainder of this paper is organized as follows: Section 2 presents the notations, assumptions, the protocol model, and the security definitions. Section 3 describes the protocol. Next, our scheme is evaluated in Section 4 based on a security analysis and a comparison with previous work. Section 5 includes concluding remarks.

## 2 Preliminary

### 2.1 Notations

Table 1 includes the notations used in the protocol description:

**Table 1.** Notations Used in This Paper

| | |
|---|---|
| $H$ | A one-way hash function, $h : \{0,1\}^* \rightarrow \{0,1\}^\lambda$ |
| $x_i^t$ | $\lambda$-bit random number generated by a tag during time period $i$ |
| $x_i^s$ | $\lambda$-bit random number generated by a server during time period $i$ |
| $x^{rand}$ | $\lambda$-bit random number generated by a server |
| $sk_i$ | $\lambda$-bit session key between a tag and a server during time period $i$ |
| $k_i$ | $\lambda$-bit random shared secret key between a tag and a server during time period $i$ |
| $SK^*$ | a tag specific master secret key, stored by a legitimate server only |
| $x_i$ | $\lambda$-bit, generated from $SK^*$ by the server during session $i$ |
| $\oplus$, $\|$ | bitwise XOR operation and concatenation of two bit strings, respectively |
| $\wr$ | dividing a bit string into two equal parts |

### 2.2 Assumptions

A tag T is not tamper-resistant. Initially, it stores the secret key $k_1$ which is updated after each authentication session. All communication between a server and a reader is assumed to be over a private and authentic channel. In this paper, we consider Reader and Server as a single entity. Therefore, we use the terms 'Server' or 'S' interchangeably in the text. The adversary cannot compromise the server. The tag is assumed to be vulnerable to repeated key exposures; specifically, we assume that up to $t<N$ periods can be compromised. Our goal is to minimize the effect such compromises will have. When a secret key is exposed, an adversary will be able to trace the tag for period i until the next single secure authentication session. Our notion of security is that this is the best an adversary can do. In particular, the adversary will be unable to trace a tag for any of the subsequent periods. It is assumed that hash and PRNG take the same amount of execution time. Splitting and concatenation operations take negligible amounts of time.

**Assumption 1**

A one-way hash function H can map an input x to an output of fixed length λ, which is called hash value h: $h = H(x)$. One-way hash function H has the following property:

- Given a hash value h, it is computationally infeasible to find the input x such that $H(x) = h$

**Assumption 2**

Let $z = H(x,y)$, where H is a one-way hash function. Given z and x, it is computationally infeasible to find y.

**Remark 1**: The one-time pad is a simple, classical form of encryption (See, [17] for discussion). Briefly stated, if two parties share a secret onetime pad p, for example a random bit string, then one party may transmit a message m secretly to the other via the ciphertext $p \oplus m$, where $\oplus$ denotes the XOR operation. It is well known that this form of encryption provides unconditional security. We use one-time pad, as it requires only the lightweight computational process of XORing.

**2.3 An Overview of KIMAP**

We make some slight twists to the original idea of key-insulation to design our protocol. RFID tags do not have the capability to communicate with the server by themselves, and do not have the resources for computing public key encryption, which requires high computation ability. Therefore, in our protocol, both the server and the tag use symmetric key encryption. We use only the hash function, random number generation and bitwise XOR operations. Our protocol uses two different values for server authentication by tag, and tag authentication by server. For server authentication, a part of the tag's secret key is used. The session key sk is used to authenticate the tag. To be precise, part of the tag's secret key is used to authenticate a valid server. Another part of the tag's secret key is used along with a part of a random number sent by the server to generate sk. The server generates this partial key from the master key (SK*) stored in its memory. A tag's secret key is updated during each time period. Thus, during a particular time period i, a tag computes the session key from the secret value stored in its memory and the partial secret key sent by the server by using simple XOR and hash functions. Both the tag authentication key (session key sk and tag secret key k are different in each session. They both are randomized and refreshed using the fresh values (generated from master secret SK*) sent by the legitimate server in each session. We define indistinguishability, forward and backward security. While in general, backward security means that all the future sessions will be secure if the current secret is exposed; restricted backward security for RFID says that the adversary must miss the necessary protocol transcripts to update the compromised key. In KIMAP, a tag computes a session secret key from a random value chosen by the server. This session secret key sk is used by the server to authenticate the tag. Then the server generates a message authentication value by using a part of the tag's secret key. The tag authenticates the server by verifying its value. The use of two different values to authenticate the tag and server ensures the backward and forward security of the tag. In other words, even if the sk or the tag secret key is exposed during an authentication session i, a tag's privacy is guaranteed for all other past time periods (forward security). All the future time periods are secure as well, assuming that the adversary misses the necessary protocol transcripts to update the compromised key (restricted backward security).

### 2.4 The Model

We design the model following the model proposed in [10]. However, our model is slightly different than that in [10]. We assume a fixed, polynomial-size tag set TS = $\{T_1,...,T_n\}$, and a server 'Server' as the elements of an RFID system. A Server has information for TS's authentication such as tag's secret key, master key, etc. Before the protocol is run for the first time, an initialization phase occurs in both $T^l$ and Server, where l = 1,..,n. That is, each $T^l \in$ TS runs an algorithm G to generate the secret key $k^l$, and Server also saves these values in a database field. A key-updating authentication scheme is a 5-tuple of poly-time algorithms (G, U*, S, U, Auth(AuthT/AuthS) such that:

G, the key generation algorithm, is a probabilistic algorithm which takes as input a security parameter $1^\lambda$, and the total number of tags n. It returns a master key SK*, and an initial shared key $k_1$ for each tag.

U*, the partial key generation algorithm, is a deterministic algorithm which takes as input an index i for a time period (throughout, we assume $1 \leq i \leq N$), the master key SK* and the secret key $k_i$ of a tag. It returns the partial secret key $x_i$ , for time period i.

S, the session key generation algorithm, is a deterministic algorithm which takes as input an index i, part of the tag's secret key $k_i'$, and a part of the partial secret key $x_i'$. It returns a shared session secret key $sk_i$ for time period i.

U, the tag key-update algorithm, is a deterministic algorithm which takes as input an index i, part of the tag's secret key $k_i''$, a part of the partial secret key $x_i''$, and a random $x_i^S$. It returns the tag's secret key $k_{i+1}$ for time period i + 1 (and erases $k_i$ , $x_i$ , $x_i^S$).

Auth(AuthT/AuthS), the authentication message verification algorithm, is a deterministic algorithm for a server (resp. tag) which takes as input AuthT(resp. AuthS). It returns 1 or the special symbol ⊥. AuthT/AuthS is as follows:

- AuthT/AuthS, the Tag (resp. Server) authentication message generation algorithm, is a probabilistic algorithm for a tag (resp. server) which takes as input a shared secret $sk_i$, a time period i, and random numbers $x_i^t$ and $x_i^S$ (or $x^{rand}$) ($k_i'$, $x_i$ , $x_i^S$ (or $x^{rand}$), and $x_i^t$ are the inputs for the server). It returns $\sigma_i'$ (resp. $\sigma_i$).

KIMAP is used as one might expect. A server begins by generating (SK*, $k_1$) ←G($1^\lambda$, n), storing SK*on a server (physically-secure device), and storing $k_1$inboth the server and the tag. At the beginning of time period i, the tag requests $x_i$ = U*(i, SK*, $k_i$ ) from the server. Using $x_i$ , and $k_i$ , the tag may compute the session secret key $sk_i$= S(i, $k_i'$, $x_i'$). This key is used to create authentication messages sent during time period i. Both the tag and server update their shared secret by $k_{i+1}$= U(i, $k_i''$,$x_i''$,$x_i^S$). After computation of $k_{i+1}$, the tag must erase $k_i$ , and $x_i$ .

## 2.5 Security Definitions

Adversary A's interaction with the RFID entities in the network is modeled by sending the following queries to an oracle O and receiving the result from O. The queries in our model follow [12] with some differences. We do not need Reply*/Execute*, since we do not consider a tag to be maintaining an internal state in our protocol. Also, we consider server and reader as a single entity. So, we do not need $Forward_1$/$Forward_2$and Auth queries. Instead, Reply, Reply' perform the tasks of $Forward_1$, $Forward_2$, respectively. They also serve the purpose of Auth(AuthT/AuthS).

--- Query(S, $x_i^S$): It calls server (S) and outputs $x_i^S$of period i.

--- Query'($T_i^l$, $x_i^t$): It calls tag ($T^l$) and outputs $x_i^t$ of period i.

--- $Query_b$ (S, $x^{rand}$): It calls server (S) and outputs any random $x^{rand}$.

--- Reply(S,$x_i^t$,$\sigma_i$,$\delta_i$): It calls S with input $x_i^t$and outputs $\sigma_i$, $\delta_i$ for period i. It uses AuthS algorithm. The output is forwarded to $T^l$.

---Reply'($T_i^l$, $x_i^s$, $\sigma_i$, $\delta_i$, $\sigma_i'$): It calls $T^l$ with input $x_i^s$, $\sigma_i$, $\delta_i$ and outputs $\sigma_i'$ for period i. It uses AuthT algorithm. The output is forwarded to S.

---$Reply_b$($T_i^l$, $x^{rand}$, $\sigma_i$, $\delta_i$ , $\sigma_i'$): It calls $T^l$ with input $x^{rand}$, $\sigma_i$, $\delta_i$ and outputs $\sigma_i'$ for period i. It uses AuthT algorithm. The output is forwarded to S.

---Execute($T_i^l$, S): This query uses the algorithms (G, U*, S, U, Auth(AuthT/AuthS)).It receives the protocol transcripts $\sigma_i$, $\delta_i$ , $\sigma_i'$, $x_i^s$, $x_i^t$, and outputs them. This models the adversary A's eavesdropping of protocol transcripts. It has the following relationships with the above queries: Execute ($T_i^l$, S) = Query(S, $x_i^S$) ∧ Query'($T_i^l$, $x_i^t$) ∧ Reply(S, $x_i^t$, $\sigma_i$, $\delta_i$) ∧ Reply'($T_i^l$, $x_i^s$, $\sigma_i$, $\delta_i$ , $\sigma_i'$).

--- $Execute_b(T_i^l$, S): This query uses the algorithms (G, U*, S, U, Auth(AuthT/AuthS)). It receives the protocol transcripts $\sigma_i$, $\delta_i$ , $\sigma_i'$, $x_i^t$, $x^{rand}$, and outputs them. This models the adversary A's eavesdropping of protocol transcripts except $x_i^s$ which is used for key update. It has the following relationship with the above queries: $Execute_b(T_i^l$, S) = $Query_b(S, x^{rand})$ ∧ Query'($T_i^l$, $x_i^t$) ∧ Reply(S, $x_i^t$, $\sigma_i$, $\delta_i$) ∧ $Reply_b(T_i^l$, $x^{rand}$, $\sigma_i$, $\delta_i$ , $\sigma_i'$).

--- RevealSecret ($T^l$, i): This query uses the algorithm U. It receives the tag's $T^l$, and outputs $k_i$ of period i.

--- Test ($T^l$, i): This query is allowed only once, at any time during A's execution. A random bit b is generated; if *b*= 1, A is given transcripts corresponding to the tag, and if *b* = 0, A receives a random value.

We now give the definitions through security games, reminiscent of classic indistinguishability in a cryptosystem security game. We follow [12] to define indistinguishability and forward security. The success of A in the games is subject to A's advantage in distinguishing whether A has received an RFID tag's real response or a random value. The next two games represent the attack games for forward security and restricted backward security, respectively.

***Definition 1.Indistinguishability***

- Phase 1: Initialization
  (1) Run algorithm G($1^\lambda$, n) → ($k^1$,..., $k^n$).
  (2) Set each tag $T^l$'s secret key as $k^l$ , where $T^l$ ϵ TS = { $T^l$,..., $T^n$}.
  (3) Save each $T^l$'s $k^l$ generated in step (1) in Server's field.
- Phase 2: Learning
  (1) $A^{ind}$ executes Query(S, $x_i^S$), Query'($T_i^l$, $x_i^t$), Reply(S, $x_i^t$, $\sigma_i$, $\delta_i$),Reply'($T_i^l$, $x_i^s$, $\sigma_i$, $\delta_i$ , $\sigma_i'$), and Execute($T_i^l$, S) oracles for all n-1 tags, except the $T^c$ ϵ TS used in challenge phase.
- Phase 3: Challenge
  (1) $A^{ind}$ selects a challenge tag $T^c$ from TS.
  (2) $A^{ind}$ executes Query(S, $x_i^S$), Query'($T_i^l$, $x_i^t$), Reply(S, $x_i^t$, $\sigma_i$, $\delta_i$),Reply'($T_i^l$, $x_i^s$, $\sigma_i$, $\delta_i$ , $\sigma_i'$), and Execute($T_i^l$, S) oracles for $T^c$, where i = 1,..., q-1.
  (3) $A^{ind}$ calls the oracle Test($T^c$, i).
  (4) For the $A^{ind}$'s Test, Oracle O tosses a fair coin *b* ϵ {0, 1}; let $b \leftarrow_R$ {0, 1}.

i. If *b*= 1, $A^{ind}$ is given the messages corresponding to $T^c$'s i-th instance.

ii. If *b*= 0, $A^{ind}$ is given random values.

(5) $A^{ind}$ outputs a guess bit *b*'.

A wins if *b*= *b*'

The advantage of any PPT adversary $A^{ind}$ with computational boundary $e_1$, $r_1$, $r_2$, λ, where $e_1$ is the number of Execute, $r_1$ is the number of Reply, $r_2$ is the number of Reply' and λ is the security parameter, is defined as follows:

$Adv_{Aind}= |Pr[b= b'] - 1/2|$

The scheme provides indistinguishability if and only if the advantage of $Adv_{Aind}$ is negligible.

***Definition 2.Forward Security***

- Phase 1: Initialization
  (1) Run algorithm $G(1^\lambda, n) \rightarrow (k^1,\ldots, k^n)$.
  (2) Set each tag $T^l$ s secret key as $k^l$ , where $T^l \in TS = \{T^l,\ldots, T^n\}$.
  (3) Save each $T^l$'s $k^l$ generated in step (1) in Server's field.
- Phase 2: Learning
  (1) $A^{for}$ executes Query(S, $x_i^s$), Query'($T_i^l$, $x_i^t$), Reply(S, $x_i^t$, $\sigma_i$, $\delta_i$),Reply'($T_i^l$, $x_i^s$, $\sigma_i$, $\delta_i$ , $\sigma_i'$), and Execute($T_i^l$, S) oracles for all n-1 tags, except the $T^c \in TS$ used in challenge phase.
- Phase 3: Challenge
  (1) $A^{for}$ selects a challenge tag $T^c$ from TS.
  (2) $A^{for}$ executes Query(S, $x_i^s$), Query'($T_i^l$, $x_i^t$), Reply(S, $x_i^t$, $\sigma_i$, $\delta_i$),Reply'($T_i^l$, $x_i^s$, $\sigma_i$, $\delta_i$ , $\sigma_i'$), and Execute($T_i^l$, S) and RevealSecret ($T_i^l$, i) oracles for $T^c$ for $T^{c}$'s i-th instance.
  (3) $A^{for}$ calls the oracle Test ($T^c$, i-1).
  (4) For the $A^{for}$'s Test, Oracle O tosses a fair coin $b \in \{0, 1\}$; let $b \leftarrow_R \{0, 1\}$.
    i. If $b= 1$, $A^{for}$ is given the messages corresponding to $T^c$'s (i-1)-th instance.
    ii. If $b= 0$, $A^{for}$ is given random values.
  (5)$A^{for}$ executes the oracles for n-1 tags of TS, except $T^c$, like in the learning phase.
  (6) $A^{for}$ outputs a guess bit $b'$.
  A wins if $b= b'$

The advantage of any PPT adversary $A^{for}$ with computational boundary $e_1$, $r_1$, $r_2$, λ, where $e_1$ is the number of Execute, $r_1$ is the number of Reply, $r_2$ is the number of Reply' and λ is the security parameter, is defined as follows:

$Adv_{Afor}= |Pr[b= b'] - 1/2|$

The scheme is forward secure if and only if the advantage of $Adv_{Afor}$ is negligible.

***Definition 3.Restricted Backward Security***

- Phase 1: Initialization
  (1) Run algorithm $G(1^\lambda, n) \rightarrow (k^1,\ldots, k^n)$.
  (2) Set each tag $T^l$ s secret key as $k^l$ , where $T^l \in TS = \{T^l,\ldots, T^n\}$.
  (3) Save each $T^l$'s $k^l$ generated in step (1) in Server's field.
- Phase 2: Learning

(1) $A^{back}$ executes $Query_b(S, x^{rand})$, $Query'(T_i^l, x_i^t)$, $Reply(S, x_i^t; \sigma_i, \delta_i)$, $Reply_b(T_i^l, x^{rand}, \sigma_i, \delta_i, \sigma_i')$, and $Execute_b(T_i^l, S)$ oracles for all n-1 tags, except the $T^c \in TS$ used in challenge phase.

- Phase 3: Challenge

(1) $A^{back}$ selects a challenge tag $T^c$ from TS.

(2) $A^{back}$ executes $Query_b(S, x^{rand})$, $Query'(T_i^l, x_i^t)$, $Reply(S, x_i^t, \sigma_i, \delta_i)$,$Reply_b(T_i^l, x^{rand}, \sigma_i, \delta_i, \sigma_i')$, $Execute_b(T_i^l, S)$ and RevealSecret $(T_i^l, i)$ oracles for $T^c$ for $T^c$'s i-th instance.

(3) $A^{back}$ calls the oracle Test ($T^c$; i+1).

(4) For the $A^{back}$'s Test, Oracle O tosses a fair coin $b \in \{0, 1\}$; let $b \leftarrow_R \{0, 1\}$

i. If $b= 1$, $A^{back}$ is given the messages corresponding to $T^c$'s (i+1)-th instance.

ii. If $b= 0$, $A^{back}$ is given random values.

(5) $A^{back}$ executes the oracles for n-1 tags of TS, except $T^c$, like in the learning phase.

(6) $A^{back}$ outputs a guess bit $b'$.

A wins if $b= b'$

The advantage of any PPT adversary $A^{back}$ with computational boundary $e_1$, $r_1$, $r_b$, $\lambda$, where $e_1$is the number of Execute, $r_1$ is the number of Reply, $r_b$ is the number of $Reply_b$ and $\lambda$ is the security parameter, is defined as follows:

$Adv_{Aback}= |Pr[b= b'] - 1/2|$

The scheme is restricted backward secure if and only if the advantage of $Adv_{Aback}$ is negligible.

*A note on restricted backward security:* Since once obtaining the tag secret by RevealSecret, $A^{back}$ takes all the power of the tag itself and thus can trace the target tag at least during the authentication immediately following the attack. In typical RFID system environments, tags and readers operate only at short communication range and for a relatively short period of time. Thus, the minimum restriction for backward security is such that the adversary misses the protocol transcripts needed to update the compromised secret key. The same restriction was applied in [26]. On the other hand, [14] claimed that there should exist some non-empty gap not accessible by the adversary between the time of a reveal query and the attack time. But this restriction was shown to be inadequate to provide backward security by [20].

***Definition 4.Privacy-Preserving Protocol***

A protocol is privacy-preserving when indistinguishability, forward security, and restricted backward security are guaranteed for any PPT adversary A with computational boundary $e_1$, $r_1$, $e_2$, $r_2$, $r_b$, $\lambda$, where $e_1$ is the number of Execute, $r_1$ is the number of Reply, $e_2$ is the number of $Execute_b$, $r_2$ is the number of Reply', $r_b$ is the number of $Reply_b$ and $\lambda$ is the security parameter.

# 3 Protocol Description

Table 2 describes the protocol building blocks, and Fig. 1 describes the authentication session. During any session i, the following steps take place between a tag and a server:

1. The server sends a random challenge $x_i^s$ to the tag.

2. The tag replies to the server with a random $x_i^t$.

3. The server splits $k_i$ into $k_i'$ and $k_i''$, and $x_i^t$ into $x_i^{s\prime}$ and $x_i^{s\prime\prime}$. It then generates $x_i$ from SK* and $k_i$ by $H_i(SK^*, k_i)$, where $H_i$ is the i-th time run for H. SK* is used to generate $x_i$ so that no other entities other than a valid server can generate $x_i$ . Even if an adversary compromises $k_i$, it cannot generate $x$ for any subsequent sessions using only that $k_i$ . $x_i^s$ is used as a random number for server authentication, and $x_i$ is used as the partial key for the present session. The server computes $\sigma_i = H(k_i' \| x_i , x_i^s \| x_i^t)$, and $\delta_i = k_i \oplus x_i$. The server sends $\sigma_i$ and $\delta_i$ to the tag.

4. After receiving $\sigma_i$ and $\delta_i$, the tag splits $k_i$ into $k_i'$ and $k_i''$, and extracts $x_i$ from $\delta_i$. The tag then authenticates the server by verifying $\sigma_i$. If the server is authenticated as a legitimate server, the tag splits $x_i^s$ into $x_i^{s\prime}$ and $x_i^{s\prime\prime}$, and $x_i$ into $x_i'$ and $x_i''$. The tag now computes the session secret key $sk_i$ by concatenating $k_i'$ and $x_i'$. It then computes $\sigma_i' = H(x_i^t \| x_i^s, sk_i)$, and updates its own secret key to $k_{i+1}$ by $H(k_i'' \| x_i'', x_i^s)$. The tag sends $\sigma_i'$ to the server, and erases $x_i$ , $x_i^t$, and $sk_i$ from its memory. The updated $k_{i+1}$ is used for the next authentication session.

5. After the server receives $\sigma_i'$, it authenticates the tag by verifying $\sigma_i'$. Theserver then updates the secret key to $k_{i+1}$ of the tag by $H(k_i'' \| x_i'', x_i^s)$. This updated $k_{i+1}$ is stored in the server database, and is used for the next authentication session.

*A note on possible timing attack:* Note that it is imperative for the respective times taken by authentication success and failure to be as close as possible to prevent obvious timing attacks by malicious readers (aimed at distinguishing among the two cases)[29]. For this reason, even if the authentication by a tag is failed, it should generate random numbers instead of simply failure, to make the cases of success and failure indistinguishable from each other.

**Table 2.** Protocol Building Blocks

| | |
|---|---|
| $\mathcal{U}^*$: | Auth (AuthT/ AuthS) |
| input: $i, SK^*, k_i$ | AuthT: |
| compute: $H_i(SK^*, k_i)$ | input: $i, x_i^t, x_i^s, sk_i$ |
| return: $x_i$ | compute: $H(x_i^t \| x_i^s, sk_i)$ |
| $\mathcal{S}$: | return: $\sigma_i'$ |
| input: $i, k_i', x_i'$ | AuthS: |
| compute: $k_i' \| x_i'$ | input: $i, x_i^s, k_i', x_i, x_i^t$ |
| return: $sk_i$ | compute: $H(k_i' \| x_i, x_i^s \| x_i^t)$ |
| $\mathcal{U}$: | return: $\sigma_i$ |
| input: $i, k_i'', x_i'', x_i^s)$ | return: 1 or $\perp$ |
| compute: $H(k_i'' \| x_i'', x_i^s)$ | |
| return: $k_{i+1}$ | |

| **Tag:** | | **Server:** |
|---|---|---|
| $k_i$ | | $SK^*, k_i$ |
| | | $x_i^s \in \{0,1\}^*$ |
| | $\xleftarrow{x_i^s}$ | |
| $x_i^t \in \{0,1\}^*$ | | |
| | $\xrightarrow{x_i^t}$ | |
| | | $k_i = k_i' \wr\wr k_i''$ |
| | | $x_i^s = x_i^{s'} \wr\wr x_i^{s''}$ |
| | | $\mathcal{U}^*(i, SK^*, k_i) \rightarrow x_i$ |
| | | $\text{AuthS}(i, k_i', x_i, x_i^s, x_i^t) \rightarrow \sigma_i$ |
| | | $\delta_i = k_i \oplus x_i$ |
| | $\xleftarrow{\sigma_i, \delta_i}$ | |
| $k_i = k_i' \wr\wr k_i''$ | | |
| $x_i = \delta_i \oplus k_i$ | | |
| Auth(AuthS)$\rightarrow$ 1 or $\perp$ | | |
| $x_i^s = x_i^{s'} \wr\wr x_i^{s''}$ | | |
| $x_i = x_i' \wr\wr x_i''$ | | |
| $\mathcal{S}(i, k_i', x_i') \rightarrow sk_i$ | | |
| $\text{AuthT}(i, x_i^t, x_i^s, sk_i) \rightarrow \sigma_i'$ | | |
| $\mathcal{U}(i, k_i'', x_i'', x_i^s) \rightarrow k_{i+1}$ | | |
| | $\xrightarrow{\sigma_i'}$ | |
| | | Auth(AuthT)$\rightarrow$ 1 or $\perp$ |
| | | $\mathcal{U}(i, k_i'', x_i'', x_i^s) \rightarrow k_{i+1}$ |

**Fig. 1.** Our Scheme: KIMAP

## 4 Evaluation

### 4.1 Security Analysis

We use a proof method similar to that described in [13, 12]. Even though the protocol in our model and that in [13, 12] are different, a similar proof can be used because both are based on the impossibility of distinguishing any two values. Before stating the theorems, we introduce the following lemma:

**Lemma 1.** Let L = x⊕y, where x, y ϵ $\{0, 1\}^k$. L is fixed. Then there exist $2^k$ pairs of (x, y) such that L = x⊕y.

**Proof :** For $\forall y$ ϵ $\{0, 1\}^k$, x is defined as x= L⊕y. We assume that ∃(x, y), (x′, y) where x≠x′, L = x⊕y, and L = x′⊕y. Then y = L⊕x = L⊕x′ ⟹x = x′. Therefore, $\forall y$ ϵ $\{0, 1\}^k$, $\exists_1$z ϵ $\{0, 1\}^k$ such that L = x⊕y. □

**Theorem 1.** The protocol π= (G, U*, S, U, Auth(AuthT/AuthS) provides indistinguishability for any PPT adversary $A^{ind}$ with computational boundary $e_1$, $r_1$, $r_2$, λ, where $e_1$ is the number of Execute, $r_1$ is the number of Reply, $r_2$ is the number of Reply' and λ is the security parameter.

**Proof:**

Let us define simulators $Sim^{exe}$, $Sim^{Que}$, $Sim^{Que'}$, $Sim^{Rep}$, and $Sim^{Rep'}$ for $T^c$ in the indistinguishability game. These simulators do not know the value of b or any secret key $k^c$ or $sk^c$ for $T^c$. $A^{ind}$'s interaction with the simulators will be computationally indistinguishable from an interaction with $T^c$. Therefore, we suppose that $A^{ind}$ does not gain knowledge from its interaction with $T^c$ in a real RFID system.

$A^{ind}$ chooses challenge tag $T^c$. Let L be the full list of real quintuplets ($rn_1$, opv, $hv_1$, $rn_2$, $hv_2$) output by $T^c$ during the challenge phase of the game, where $hv_a$ means a hashed value for a = 1, 2, $rn_b$ is a random number for b= 1, 2 and opv means one-time pad value. During the challenge phase, $Sim^{exe}$ simulates the result of an Execute call to $T^c$ by generating quintuplet ($x_i^s$, $\delta_i^*=k_i \oplus x_i$, $\sigma_i^*=$ H($k_i'||x_i$, $x_i^s||x_i^t$), $x_i^t$, $\delta_i^{*\prime}$=H($x_i^t||x_i^s$, $sk_i$)) for i ≤ # Execute= $e_1$ and appending it to a list L′.

L′ is empty at the beginning of the challenge phase and N-1 = $e_1$, where N-1 indicates the maximum number of queries executed by $A^{ind}$ for $T^c$'s instance. In addition to any valid tag quintuplets output by $Sim^{exe}$, Server includes any quintuplet in L′. In order for $A^{ind}$ to distinguish between the simulated challenge phase and a real challenge phase, $A^{ind}$ must be able to determine that some quintuplet ($x^s$, $\sigma^*, \delta^*$, $x^t$, $\sigma^{*'}$) ϵ L is invalid for $T^c$. To determine this, $A^{ind}$ must identify a quintuplet ($x^s$, $\sigma^*$ =H($k'$ $\|k$ , $x^s \|x^t$ ,$\delta^*$ =$k$ $\oplus x$ , $x^t$ , $\sigma^{*\prime}$ =H($x^t$ $\|x^s$ , sk)) that is valid for $T^c$, but such that, σ ≠$\sigma^*$,$\delta$≠$\delta^*$, and σ ≠$\sigma^{*'}$. That is, $A^{ind}$ has to remove an invalid ($x^s$ , $\delta^*$, $\sigma^*$, $x^t$ , $\sigma^{*'}$) from L′ to show that the correct $Sim^{exe}$ is present.

At some point in the course of the challenge phase, one of the following cases must occur:

1. There is a random pair ($x^s$, $x^t$) s.t. ($x^s$, $\delta^*$, $\sigma^*$, $x^t$ , $\sigma^{*'}$) ϵ L′ and ($x^s$, $\delta$, σ, $x^t$, $\sigma'$) ϵ L for some pair (X,Y), where X=($\delta^*$, $\sigma^*$,$\sigma^{*'}$) ϵ L′, Y= ($\delta$, σ, $\sigma'$) ϵ L and σ ≠$\sigma^*$, $\delta$≠$\delta^*$, and σ ≠$\sigma^{*'}$. Since $A^{ind}$may make at most $e_1$Execute calls to $T^c$, we haveMin(#Execute, |L|)= $e_1$, where #Execute= $e_1$and |L|= N -1. As $x^s$, $x^t$are random λ-bit values, and thus the space of random numbers is $2^{2\lambda}$, it follows that this condition occurs with probability at most $e_1^2$=$2^{2\lambda}$.
2. For a pair ($x^s$, $x^t$) ϵ L′,L, $A^{ind}$directly computes $\delta$, σ, $\sigma'$that are equal to Xor Y . There exist $2^\lambda$pairs (y, y′) such that $\alpha_i$= y⊕y′from Lemma 1.Let (y, y′) ϵ {0, 1}$^\lambda$×{0, 1}$^\lambda$be the random values guessed by $A^{ind}$as chosensecrets. Therefore, the success probability of $A^{ind}$to choose (x, k) pair from $\delta$is ≃$e_1$2/$2^\lambda$ = $e_1/2^{\lambda-1}$, which is negligible given that $e_1$Execute queries are called, and $e_1$is small compared to $2^{\lambda-1}$.

Moreover, since $\sigma^*$= H(k′||x, $x^s$||$x^t$), and $\sigma^{*'}$= H($x^t$||$x^s$, sk) = H($x^t$||$x^s$, k′||x′),$A^{ind}$first must be able to find out k′or x. At this time, the probability of recovering k′ from H(k′||x; $x^s$||$x^t$) is approximately $e_1$=$2^{\lambda/2}$provided that $e_1$is small compared to $2^{\lambda/2}$, and given that $e_1$Execute queries are called. Probability of randomly choosing the other part k′′is $e_1/2^{\lambda/2}$. So, the probability to guess the correct kis $e_1^2/2^\lambda$. Again, the probability of recovering x′ or k0from H($x^t$||$x^s$, sk) = H($x^t$||$x^s$; k′||x′) is ≃ $e_1/2^{\lambda/2}$. So, the probability of guessing x or k is $e_1^2/2^\lambda$.

Therefore, $A^{ind}$ can distinguish $Sim^{exe}$ from $T^c$ with probability at most $e_1^2/2^{2\lambda}$+ $e_1/2^{\lambda-1}$+ $e_1^2/2^\lambda$, which is negligible for polynomially bounded $A^{ind}$.

Now, during the challenge phase, $Sim^{Que}$, $Sim^{Que'}$ simulate the result of Query and Query' calls to $T^c$ and S by generating random numbers $\tilde{x}_i^s$, $\tilde{x}_i^t$for i≤ #Query=#Query'= q-1 and appending them to a list $M_0$.

$Sim^{Rep}$, $Sim^{Rep'}$ simulate the result of a Reply and Reply' call to $T^c$, respectively. While $Sim^{Rep}$ generates ($\tilde{\sigma}_J$ = H(k′$_j$||$x_j$, $\tilde{x}_i^s \| \tilde{x}_i^t$), $\tilde{\delta}_j$= $x_j$⊕$k_j$) for j ≤#Reply=$r_1$=q -1and appends it to a list $M_1$, $Sim^{Rep'}$ generates ($\tilde{\sigma}'_m$=H($sk_m$; $\tilde{x}_m^s k \tilde{x}_m^t$)) for m ≤#Reply'=$r_2$=q -1 and appends it to a list $M_2$. To simplify the analysis, here we assume that the results of $Sim^{Que}$, $Sim^{Que'}$ influence the simulated results of $Sim^{Rep}$ and $Sim^{Rep'}$.This is because $\tilde{x}_i^s$,$\tilde{x}_i^t$output by $Sim^{Que}$, $Sim^{Que'}$ are included in the computation of $\tilde{\sigma}$= H(k′||x, $\tilde{x}^s$ $\|\tilde{x}^t$ ),$\tilde{\sigma}'$ = H(sk, $\tilde{x}^s$ $\|\tilde{x}^t$ ) of $Sim^{Rep}$ and $Sim^{Rep'}$, where $\tilde{x}_i^s$= $x^s$ and $\tilde{x}_i^t$= $x^t$. Of course, we can consider that the random numbers are independent in σ, σ′ which causes the complicated analysis.

Recall that $A^{ind}$ selects the challenge tag $T^c$ from the unrevealed tags, and L is the full list of quintuplets ($rn_1$, opv, $hv_1$, $rn_2$, $hv_2$) output by $T^c$ during the challenge phase of the game. Note that $M_0$,$M_1$,$M_2$are empty at the beginning of the challenge phase. In order for $A^{ind}$ to distinguish between the simulated challenge phase and a real phase, $A^{ind}$ must determine that some quadruplet ($\tilde{x}^s$ , $\tilde{\sigma}$, $\tilde{\delta}$, $\tilde{x}^t$ ) ϵ $M_1$is invalid for $T^c$. To determine this, $A^{ind}$ must identify a quadruplet ($x^s$,$\delta$= k⊕x,$\sigma$= H(k′||x, $x^s$||$x^t$), $x^t$) that is valid for $T^c$, but such that $\delta \neq \tilde{\delta}$ and $\sigma \neq \tilde{\sigma}$, to show that $Sim^{Rep}$ is present.

Consequently, one of the following two cases must occur at some point in the course of the challenge phase of the game.

1. There are random numbers $x^s$,$x^t$ such that ($\tilde{x}^s$ , $\tilde{\sigma}$, $\tilde{\delta}$, $\tilde{x}^t$ ) ϵ $M_1$and ($x^s$, δ, σ, $x^t$) ϵ L for some pair (X, Y ), where X = ($\tilde{\sigma}$, $\tilde{\delta}$) ϵ $M_1$and Y = (δ, σ) ϵ L: Since $A^{ind}$ may execute at most $r_1$Reply calls to $T^c$, we have Min(#Reply; |L|) = $r_1$, where #Reply= $r_1$and |L|= q -1. As $x^s$, $x^t$ are random λ-bit values, and thus the space of random numbers is $2^{2\lambda}$, it follows that this case occurs with probability at most $r^2{}_1/2^{2\lambda}$.
2. For random numbers $x^s$,$x^t$ ϵ $M_1$,L, $A^{ind}$ computes the values corresponding to X or Y : Since σ= H(k′||x, $x^s$||$x^t$), $A^{ind}$ must know (x, k) and ($x^s$, $x^t$) to compute σ corresponding to X or Y . Given that at most $r_1$Replyqueries are called, the probability of recovering (x, k) is $r_1/2^\lambda$. Again, for a pair ($x^s$, $x^t$) ϵ $M_1$,L, $A^{ind}$ directly computes δ that is equal to X or Y . There exist $2^\lambda$pairs (y, y′) such that $\alpha_i$= y⊕y′from Lemma 1. Let (y, y′) ϵ $\{0, 1\}^\lambda \times \{0, 1\}^\lambda$ be the random values guessed by $A^{ind}$ as chosen secrets. Therefore, the success probability of $A^{ind}$ to choose (x, k) pair from δ is≃ $r_1 2/2^\lambda$= $r_1/2^{\lambda-1}$, which is negligible given that $r_1$Reply queries are called, and $r_1$is small compared to $2^{\lambda-1}$.

Therefore, $A^{ind}$ can distinguish $Sim^{Rep}$ from $T^c$ with probability at most $r^2{}_1/2^\lambda$+ $r_1/2^\lambda$+ $r_1/2^{\lambda-1}$, which is negligible for polynomial bounded $A^{ind}$.

Similarly, $A^{ind}$ can distinguish $Sim^{Rep'}$ from $T^c$ with probability at most $r^2{}_2/2^\lambda$+ $r_2/2^\lambda$, which is negligible for polynomial bounded $A^{ind}$. ■

**Theorem 2.**The protocol π= (G, U*, S, U, Auth(AuthT/ AuthS) is forward secure for any PPT adversary $A^{for}$ with computational boundary $e_1$, $r_1$, $r_2$, λ, where $e_1$is the number of Execute, $r_1$is the number of Reply, $r_2$is the number of Reply' and λ is the security parameter.

**Proof:**

In the challenge phase, $A^{for}$ makes $e_1$Execute calls for each tag of (n -1) tags except $T^c$, as in the learning phase. At this time, let $A^{for}$'s advantage for recovering $T^i$'s secret key $k_i$ be $Adv^{Rec}{}_{Afor,Ti}(\lambda)$ from the collected transactions of Execute queries. In other words, the probability of finding out $k_i$ from a quintuplet ($x^{si}$,$\delta^i$= $k^i$⊕$x^i$,$\sigma^i$= H($k^{i'}$||$x^i$, $x^{si}$||$x^{ti}$), $x^{ti}$,$\sigma^{i'}$= H($x^{ti}$||$x^{si}$, $sk^i$)) is $1-(1-1/2^\lambda)^{e1}$,given $e_1$ Execute queries for $T^i$, and i = 1,…, n -1.

When $A^{for}$ is given a random value or $T^c$'s real message in response to Test query, it must be able to compute ($\delta^c$, $\sigma^c$, $\sigma^{c'}$) corresponding to $T^c$'s (i -1)-th instance to guess correctly, i.e. b≠b′, where $\delta^c$= $k^c$⊕$x^c$, $\sigma^c$= H($k^{c'}$||$x^c$, $x^{sc}$||$x^{tc}$), $\sigma^{c'}$=H($x^{tc}$||$x^{sc}$, $sk^c$). As the necessary condition, $A^{for}$ has to recover $k^c$(of (i -1)-th instance) from i-th instance H(k′′||x′′, $x^s$). Note that, $A^{for}$ already knows k related to the i-th instance with RevealSecret. We now define $A^{for}$'s advantage in picking k′′ by guessing the correct fair coin bas $Adv_{for}(\lambda/2)$, thus the following is induced: $Adv_{for}(\lambda/2) \leq Adv^{Rec}_{Afor,T1}(\lambda/2) + Adv^{Rec}_{Afor,T2}(\lambda/2) + \ldots + Adv^{Rec}_{Afor,Tn-1}(\lambda/2) \leq (n-1)Adv^{Rec}_{Afor;T1}(\lambda/2) \leq (n-1)\{1-(1-1/2^{\lambda/2})^{e1}\} \simeq (n-1)e_1/2^{\lambda/2}$. Again the probability of choosing k′ is $\simeq e_1/2^{\lambda/2}$. In total, $A_{for}$'s advantage in guessing the correct fair coin b is at most $(n-1)e^2{}_1/2^{\lambda}$, which is negligible.

With a similar method, we can show that forward security for $A^{for}$ is satisfied in KIMAP. When the maximum number of Reply and Reply' for (n -1)'s $T^c$ is $r_1$and $r_2$, respectively, the adversary $A^{for}$ can correctly guess b with probability at most $(n-1)\cdot r_1/2^{\lambda/2}\cdot r_1/2^{\lambda/2}+(n-1)\cdot r_2/2^{\lambda/2}\cdot r_2/2^{\lambda/2}+(n-1)\cdot r_1/2^{\lambda-1}=(n-1)\cdot r^2{}_1/2^{\lambda}+(n-1)\cdot r^2{}_2/2^{\lambda}+r_1/2^{\lambda-1}$, which is negligible for polynomial bounded $A^{for}$.■

**Theorem 3.**The protocol π= (G, U*, S, U, Auth(AuthT/ AuthS) is restricted backward secure for any PPT adversary $A^{back}$ with computational boundary $e_2$, $r_1$, $r_b$, λ, where $e_2$is the number of $Execute_b$, $r_1$is the number of Reply, $r_b$ is the number of $Reply_b$ and λ is the security parameter.

**Proof:**

Assuming that $A^{back}$ misses the necessary protocol transcripts to update the compromised key, the proof for backward security is similar to that of Theorem 2 with the exception that the future secrets are secure.

In the challenge phase, $A^{back}$ makes $e_1 Execute_b$calls for each of (n -1)'s tags except $T^c$, as in the learning phase. At this time, let $A^{back}$'s advantage in recovering $T^i$'s secret key $k^i$ be $Adv^{Rec}_{Aback,Ti}(\lambda)$ from the collected transactions of $Execute_b$ queries. In other words, the probability of finding out $k^i$ from a quintuplet $(x^{si}, \delta^i= k^i \oplus x^i, \sigma^i= H(k^{i'}||x^i, x^{si}||x^{ti}), x^{ti}, \sigma^{i'} = H(x^{ti}||x^{si}, sk^i))$ is $1-(1-1/2^{\lambda})^{e2}$, given $e_2 Execute_b$queries for $T^i$, and i = 1,..., n -1.

When $A^{back}$ is given a random value or $T^c$'s real message in response to Test query, it must be able to compute $(\delta^c, \sigma^c, \sigma^{c'})$ corresponding to $T^c$'s (i+1)-th instance to guess correctly, i.e. b≠b′, where $\delta^c= k^c \oplus x^c$, $\sigma^c= H(k^{c'}||x^c, x^{sc}||x^{tc})$, $\sigma^{c'} =H(x^{tc}||x^{sc}, sk^c)$. As the necessary condition, $A^{for}$ has to recover $k^c$ (of (i -1)-th instance) from i-th instance $H(k''||x'', x^s)$. Note that, $A^{back}$ already knows k related to the i-th instance with RevealSecret but missed the necessary transcript $x_i^s$which is used to update k for (i +1)-th instance. We now define $A^{back}$'s advantage in guessing the correct fair coin b as $Adv_{back}(\lambda)$, thus the following is induced: $Adv_{back}(\lambda) \leq Adv^{Rec}_{Aback,T1}(\lambda) + Adv^{Rec}_{Aback,T2}(\lambda) + \ldots + Adv^{Rec}_{Aback,Tn-1}(\lambda) \leq (n-1) Adv^{Rec}_{Aback,T1}(\lambda) \leq (n-1)\{1-(1-1/2^{\lambda})^{e1}\} \simeq (n-1)e_1/2^{\lambda}$, which is negligible.

Similarly, we can show that backward security for $A^{back}$ is satisfied in KIMAP. When the maximum number of Reply and $Reply_b$ for (n -1)'s $T^c$is $r_1$and $r_b$, respectively, the adversary $A^{back}$ can correctly compute $k_{i+1}$ from Reply with probability at most $(n-1)\cdot r^2{}_1/2^{\lambda}+(n-1)\cdot r_1/2^{\lambda-1}$,and from $Reply_b$ with probability at most $(n-1)\cdot r^2{}_b/2^{\lambda}$ . So, $A^{back}$ can guess b with probability at most $(n-1)\cdot r^2{}_1/2^{\lambda}+(n-1)\cdot r_1/2^{\lambda-1}+(n-1)\cdot r^2{}_b/2^{\lambda}$, which is negligible for polynomial bounded $A^{back}$.■

**Theorem 4.**The protocol π= (G, U*, S, U, Auth(AuthT/ AuthS) is privacy-preserving for any PPT adversary A with computational boundary $e_1$, $e_2$, $r_1$, $r_2$, $r_b$, λ, where $e_1$ is the number of Execute, $e_2$ is the number of $Execute_b$, $r_1$ is the number of Reply, $r_2$ is the number of Reply', $r_b$ is the number of $Reply_b$ and λ is the security parameter.

### 4.2 Discussion and Comparison with Previous Work

Deursen et al. [33] discussed a weakness of the indistinguishability definition of [12]. Deursen et al. argued that, to achieve location privacy, the adversary must not be able to distinguish one tag's response from other tags' responses, but it is not necessary that the adversary cannot distinguish the tag's response from any arbitrary value. However, our definition can be modified according to their argument. For that purpose, the oracle queries should run on all but two tags which are used for the challenge phase. All the adversary needs to do is to distinguish between those two tags. In fact, our assumption about the tag responses is such that the outputs of the one-way hash functions are indistinguishable from a random bit string of equal length.

In [3], Bellare et al. show that it is impossible to achieve public-channel key insulated security in the face of an active adversary (who can compromise the secret key). Although we follow the idea of key insulation from [10], assuming passive adversary in case of RFID (who can eavesdrop only) is not practical, as it is easy for an adversary to break into a tag's memory. Considering this, the assumptions made in our scheme (as well as in [26]) are more realistic to achieve restricted backward security and the other features as well. However, many of the existing mutual authentication protocols may support restricted backward security under our assumption ([5, 31, 29] to name a few). But [5, 31] require a tag to remember too many secrets. Moreover, [5, 31] cannot provide forward security as shown by [23] and [35], respectively. Again, [29] requires more computation than our scheme, and it does not provide reader authentication. Nevertheless, none of these protocols came up with a formal model of backward security (even if not perfect).

Although it is not the primary target of our proposed protocol, it is also possible to prevent desynchronization attacks [32] in our protocol to some extent. We consider the following type of attack: If the last message is blocked, the tag updates the shared secret key, $k_i$, but the server doesn't. The server and tag are no longer able to communicate successfully. To prevent such an attack, the server has to remember the last valid authentication session transcripts and the secret values. When a server receives some random number instead of a valid authentication value from a tag, the server updates itself using the information from the last valid session, and tries again to get synchronized with the tag. Although the question of scalability is an issue here, this approach can help avoid such desynchronization attacks in a limited way (of course the system gets desynchronized if the last messages from two consecutive sessions are blocked). Even though the system gets desynchronized, an adversary cannot trace a tag from its desynchronized state, since the responses of a tag are always pseudorandom, hence indistinguishable. However, as discussed in [8], eliminating any possibility of desynchronization is difficult on a technical level given the limited functionality of low-cost tags. Nevertheless, providing tools for detecting such an attack and localizing the adversarial device is not a major issue. In actual systems, the operator would have to physically remove or deactivate the attack device. In this paper, we are more concerned with `exposure resilience' of the secret key and its effect on the authentication protocol, rather than the desynchronization attacks. Providing full resistance against desynchronization attacks is a separate issue.

Unlike other work mentioned in Section 1.1, we achieve both forward and backward security, along with indistinguishability. The keys used for tag authentication in the protocol are different in each session for each tag. Unlike MSW protocols, it is not possible in our protocol for an adversary to derive secrets of other tags, even if she obtains a tag's secret key since no two tags share any secret key. We compare our work, based on security properties and computational cost, with LK and SM schemes in Table 3 below. According to [12], a scheme must satisfy both forward security and indistinguishability in order to achieve `strong location privacy'. If a scheme satisfies indistinguishability only, the scheme is `weak location private'. [25] has shown that SM scheme is not forward secure. So, SM scheme is weak location private only, whereas our scheme is strong location private. SM scheme furthermore does not give any formal security model for indistinguishability and forward security. Regarding computational requirements, our protocol requires a simple one-way hash function, random number generation and the XOR operation. We use a simple hash function like DM-PRESENT-80 [4] to achieve forward security for the tag. This requires around 2200 gates.

Assuming that generating random numbers requires the same computational ability as for the hash functions, our protocol needs 4 hash operations only. According to Table 3, in KIMAP, a tag requires fewer computations compared to the LK and SM schemes. Assuming that a secret key is 64-bits long, a tag in the LK scheme requires storing 3 secret keys (192 bits). As the server needs to authenticate itself first to a tag, the server must broadcast the authentication messages to the tags. Since the server does not know the id of the tag that it wants to authenticate, the server has to compute and broadcast the authentication messages for all the tags in its storage. We assume that the server has enough resource to perform such computation. On the other hand, a tag receiving the broadcast messages has to find a match with its verification value. Although computing the verification value is always constant, finding a match increases the required computations according to the number of broadcast messages in the worst case. As stated earlier, such a scenario is unavoidable when we require that a server should authenticate itself first to a tag. We say that our scheme is more suitable for an environment where the reader must read a number of tags at a time (inventory management) and/or where there are not too many tags (library with a few thousand books).

Table 3: Performance Comparison

| Feature / Schemes | Ind. | Forward Security | Restricted Backward Security | Tag's computation | Tag's Storage* |
|---|---|---|---|---|---|
| LK'06 | OK | OK | X | 2 XOR, 5 Hash | 384 bits |
| SM'08 | OK | X | OK | 6 XOR, 4 Hash | 128 bits |
| Our Work | OK | OK | OK | 1 XOR, 4 Hash | 128 bits |

Ind.: Indistinguishability; OK: Scheme Satisfies The Feature; X: Scheme Does Not Satisfy The Feature; XOR: Bitwise XOR operation; * Each secret is 128 bits

**4.3 A Quantitative Performance Analysis**

In most of the RFID applications, there is a tag reader between tags and the server. For the performance analysis, we assume that a reader exists, and it only transfers the message transcripts to tag and server. Before making our assumptions and analyzing our schemes, we briefly review some facts on practical deployment of an RFID system.

- Tag readers are assumed to have a secure and dedicated connection to a back-end database. Although readers in practice may only read tags from within the short tag operating range, the reader-to-tag, or forward channel is assumed to be broadcast with a signal strong enough to monitor from long range. The tag-to-reader or backward channel is relatively much weaker, and may only be monitored by eavesdroppers within the tag's shorter operating range [34].
- The time available for a complete reading/authentication procedure is in the range of 5-10 milliseconds considering the performance criteria of an RFID system that demands a minimum tag reading speed of at least 200 tags per second [8].
- In accordance with EPC C1G2 protocol, a maximum tag-to-reader data transmission rate bound of 640 kbps and a reader-to-tag data transmission rate bound of 126 kbps [8].
- In the low-cost tags, the complexity of implementing robust PRNGs is equivalent to the complexity of implementing robust one-way hash functions. The same assumption has been widely considered in cryptographic literature, [29,17,28] to name a few.

Our Assumptions: Based on the above facts, we use the following assumptions for the quantitative analysis of our schemes.

- The tag reading speed is at least 200 tags per second.
- The time for a complete reading/authentication procedure is in the range of5-10 milliseconds.

- The tag-to-reader data transmission rate bound is 640 kbps and a reader-to-tag data transmission rate bound is 126 kbps.
- Computing a PRNG and a one-way hash function takes same time.
- DM-PRESENT-80 hash function is used as the underlying one-way hash function. It provides 64-bit security level, and operates in a single block with 33 cycles per block at the rate of 100 khz. Each of the hash function takes 33/100khz = 0.33 milliseconds to run on a tag.
- The time required for XOR and concatenation operations is ignored, since they take negligible amount of time and resource.
- Computation time in reader and the server is ignored since reader and server have ample computational power.

Quantitative Performance: Our protocol requires 4 hash operations for a tag, thus taking 1.32 milliseconds. The tag-to-reader communication requires128-bits, so it will take 128bits/640kbps= 0.20 milliseconds. Similarly, the reader-to-tag communication takes 1.52 milliseconds. In total, our estimated total protocol execution time is: (1.32 + 0.20 + 1.52) ≈3.0 milliseconds, which is well within the requirement as stated above. Assuming that 200 tags are read at one time in case of batch-mode environment (i.e., where many tags are authenticated at once), the estimated run times of our protocols are well within the bound. As for the required number of gates, our protocols are also well within the requirements for the low-cost tags, which are expected to have 2000-5000 gates available for security purposes [8].

Most RFID readers have serial interfaces using RS/EIA 232 standards (point to point, twisted pair)[11]. Readers communicate with the back-end server using such an interface. RS 232 serial interface standard says that the bit rate is lower than 20,000 bits per second [27]. As per our assumption, when 200 tags are read at a time in the batch-mode, it would take 25600-bits to be transferred, requiring around 1.28 sec to transfer the data. In the case of reading a single tag, the data transfer would take 128bits/20000bps = 6.40 milliseconds. This amount of time for protocol execution is well within the capability of today's RFID systems.

## 5 Conclusion

We have proposed KIMAP, a privacy-preserving mutual RFID authentication protocol for shared key environment. The protocol uses two different keys for mutual authentication. The server sends a random partial key (generated from a master secret key SK*) to a tag. The tag generates the session key sk to authenticate itself to the server. The tag's secret key k is updated using a partial key received from the server. As k is purely fresh for every time period, the tag's security is guaranteed for all other time periods (both for the past and future) under our assumptions. We show that our scheme is computationally more efficient than the SM and LK schemes. Our protocol satisfies indistinguishability, and achieves both forward and restricted backward security through key-insulation. We provide a formal security model of the proposed protocol as well.